\documentclass[aps,prl,preprint,groupedaddress]{revtex4}
\usepackage{graphicx}
\usepackage{dcolumn}
\usepackage{bm}
\usepackage{comment}

\begin{document}


\title{Incommensurate Chiral CDW in 1\textit{T}-VSe$_{2}$}%

\author{Y. Sugawara$^{1}$}
\author{A. Nomura$^{1}$}
\author{Y. Toda$^{1}$}
\author{T. Kurosawa$^{2}$}
\author{M. Oda$^{2}$}
\author{T. Matsuura$^{1}$}
\author{K. Ichimura$^{1}$}
\author{S. Tanda$^{1}$}
 \affiliation{$^{1}$Department of Applied Physics, Hokkaido Univesity}
 \affiliation{$^{2}$Department of Physics, Hokkaido University}


\date{\today}

\setlength{\baselineskip}{40pt}

\maketitle

\section{abstract}
We have investigated the chiral charge-density wave (CDW) in 1\textit{T}-VSe$_{2}$ using scanning tunneling microscopy (STM) measurements and 
optical polarimetry measurements.
With the STM mesurements, we revealed that the CDW intensities along each triple-$q$ directions are different.
Thus the rotational symmetry of 1\textit{T}-VSe$_{2}$ is lower than that in typical two-dimentional triple-$q$ CDWs.
We found that the CDW peaks form a kagome lattice rather than a triangular lattice. 
The Friedel oscillations have the chirality and the periodicity reflected properties of the background CDW.
With the optical measurements in 1\textit{T}-VSe$_{2}$, we also observed a lower rotational symmetry with the polarization dependence of the transient reflectivity variation, which is consistent with the STM result on a microscopic scale.
Both 1\textit{T}-TiSe$_{2}$ and 1\textit{T}-VSe$_{2}$ show chiral CDWs, which implies that such waves are usual for CDWs with the condition 
$H_\mathrm{CDW} \equiv q_{1}\cdot(q_{2} \times q_{3}) \neq0$.

\section{INTRODUCTION}

The concept of chirality has recently been receiving a lot of attention in many fields,
including chiral condensates and chiral density waves \cite{chiraldw} in quantum chromodynamics, 
chiral spin texture \cite{texture} and electrical magnetochiral anisotropy \cite{emcha1,emcha2,emcha3} in condensed matter physics,
cholesteric liquid crystal and chiral helimagnetic order in application fields, and more.
In relation to superconductors, chiral $p$-wave superconductivity \cite{pwave} is also attracting attention.
In particular, it is suggested that the pseudogap state in a high temperature superconductor is related to chiral order \cite{hightc1,hightc2} 
and this property must be understood if we are to clarify the high-$T\rm_{c}$ supercondutor mechanism. Thus chirality is related to many fields, from elementary particles to macroscopic phenomena.

In recent years, the chiral charge-density wave (CDW), which is a new class of two-dimensional triple-$q$ CDW, was discovered in 1\textit{T}-TiSe$_{2}$ \cite{ishi1, ishi2}.
In transition metal dichalcogenides MX$_{2}$ such as 1\textit{T}-TaS$_{2}$, which are well known as typical two-dimensional CDW materials \cite{MX2},
the intensities of CDW wave vectors $q_{1}$, $q_{2}$ and $q_{3}$ have the same magnitude. 
However in chiral CDWs, these values are different.
Hence, the structure of the charge distribution has a clockwise or anticlockwise intensity anisotropy, so that the symmetry in a chiral CDW is less than in other MX$_{2}$ CDWs.
In other words, it is a system that is broken in terms of both inversion and rotational symmetry.
This phenomenon has attracted a lot of attention as regards similarities between the high-$T\rm_{c}$ pseudogap state and charge-parity symmetry.
A chiral CDW is believe to be a widespread phenomenon because it has also been observed in 2\textit{H}-TaS$_{2}$ \cite{2h}, 
but its characteristics are as yet known.
$H_\mathrm{CDW}$ is cited as the condition for CDW chirality and is defined as 
$H_\mathrm{CDW} \equiv q_{1}\cdot(q_{2} \times q_{3})$, where $q_{1}$, $q_{2}$ and $q_{3}$ are the CDW $q$ vectors 
or nesting vectors with $c^*$ components in triple-$q$ systems \cite{ishi1}.
However, as yet there are no materials that fulfill the condition and can be confirmed to be a chiral CDW except for 1\textit{T}-TiSe$_{2}$.
Therefore it is necessary to elucidate the condition of the chiral CDW.

We note that the new material 1\textit{T}-VSe$_{2}$ offers the potential for a chiral CDW.
It is unique among MX$_{2}$ CDW materials because an X-ray diffraction measurement \cite{xray} confirmed that
its CDW is commensurate in two-dimensional layers but incommensurate in the direction perpendicular to the layers.
The CDW wave vector in the $c^*$ direction is 0.314$c^*$ above 85 K and 0.307$c^*$ below 85 K \cite{xray2}. 
These values are quite different from the commensurate value of 0.333$c^*$.
This difference cannot be explained from the mechanism whereby most of the MX$_{2}$ undergoes commensurate CDW in the interlayer direction 
by stacking to avoid the CDW peaks of each layer overlapping and thus benefit the Coulomb energy.
Thus, with 1\textit{T}-VSe$_{2}$, CDWs may be commensurate both in the layer direction and the interlayer directions.
The Fermi surface of 1\textit{T}-VSe$_{2}$ obtained by band calculations \cite{band} and angle resolved photoemission spectroscopy \cite{arpes, arpes2} suggests that 
its nesting vectors have $c^*$ components.
So its CDW is three-dimensional although 1\textit{T}-VSe$_{2}$ is a two-dimensional layered material.

In this work, we investigated 1\textit{T}-VSe$_{2}$ as regards the chirality.
We found the chirality in the 1\textit{T}-VSe$_{2}$ CDW by using scanning tunneling microscopy (STM) and optical polarimetry measurements.
In the STM measurements, we found different intensity ratios for three CDW directions and the CDW peaks formed a kagome lattice.
In optical polarimetry measurements, we found a lower two-fold symmetry in 1\textit{T}-VSe$_{2}$ 
than in a typical two-dimensional triple-$q$ CDW such as 1\textit{T}-TaS$_{2}$.
By analyzing STM images, we found that Friedel oscillations on chiral CDWs reflect the chairality and periodicity of the underlying CDW.

\section{EXPERIMENTAL}

The samples were grown by the chemical vapor transport method.
The elements (vanadium rod and selenium flake) in evacuated silica tubes were heated at 750 Ž for 5 days. 
We obtained single crystals of 1\textit{T}-VSe$_{2}$, which were platelets as large as 2~2~0.1 mm$^{3}$.

In the STM measurements, the samples were cleaved in situ just before the STM tip approached the surface in an ultra high vacuum at 77 K 
and we obtained samples with clean surfaces.
The STM images were obtained at 80 K and 8 K.

The pump-probe pulse method was employed with the sample surface using the micro-optics at 4 K.
The pump and probe pulse polarization was varied to investigate the dominant direction of the deviation of the reflectivity $\Delta$R(t).

\section{RESULTS AND DISCUSSION}

\begin{figure}
\includegraphics{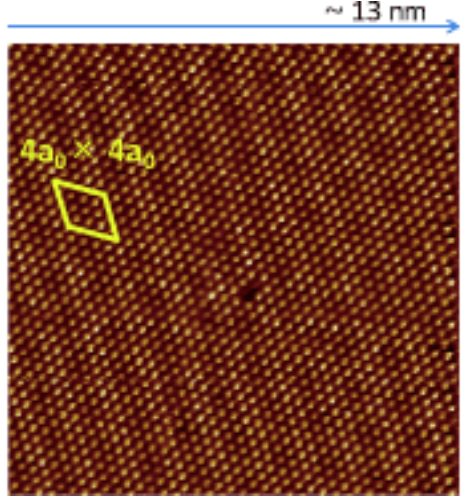}
\caption{\label{stm}13 nm ~ 13 nm STM image measured in situ at a sample voltage of $V$  -100 mV and an initial tunneling current of $I$  2.0 nA 
at 80 K. The spots correspond to Se atoms and the brightness corresponds to local electron density. 
The orange parallelogram indicates the 4$a_{0}$ ~ 4$a_{0}$ CDW superlattice.}
\end{figure}

Figure \ref{stm} shows an STM current image obtained at the $a$-$b$ plane of 1\textit{T}-VSe$_{2}$ at 80 K.
The CDW transition temperature of 1\textit{T}-VSe$_{2}$ is 110 K \cite{vse3}, so this image shows the electronic property below the CDW transition temperature.
The spots correspond to Se atoms and the brightness represents the magnitude of the local electron density.
We can observe the periodic modulataion of the electron density over the entire image.
It has a 4$a_{0}$~4$a_{0}$ periodicity indicated by the orange parallelogram in Fig. \ref{stm}.
This periodicity corresponds to the 4$a_{0}$ CDW superlattice, which is consistent with a report that employed X-ray diffraction measurements\cite{xray2}.
However, compared with a typical two-dimensional triple-$q$ CDW, such as 1\textit{T}-TaS$_{2}$, this CDW structure appears to be different because 
there is a difference in brightness that corresponds to the CDW intensity in each CDW direction.

\begin{figure}
\includegraphics[width=12cm]{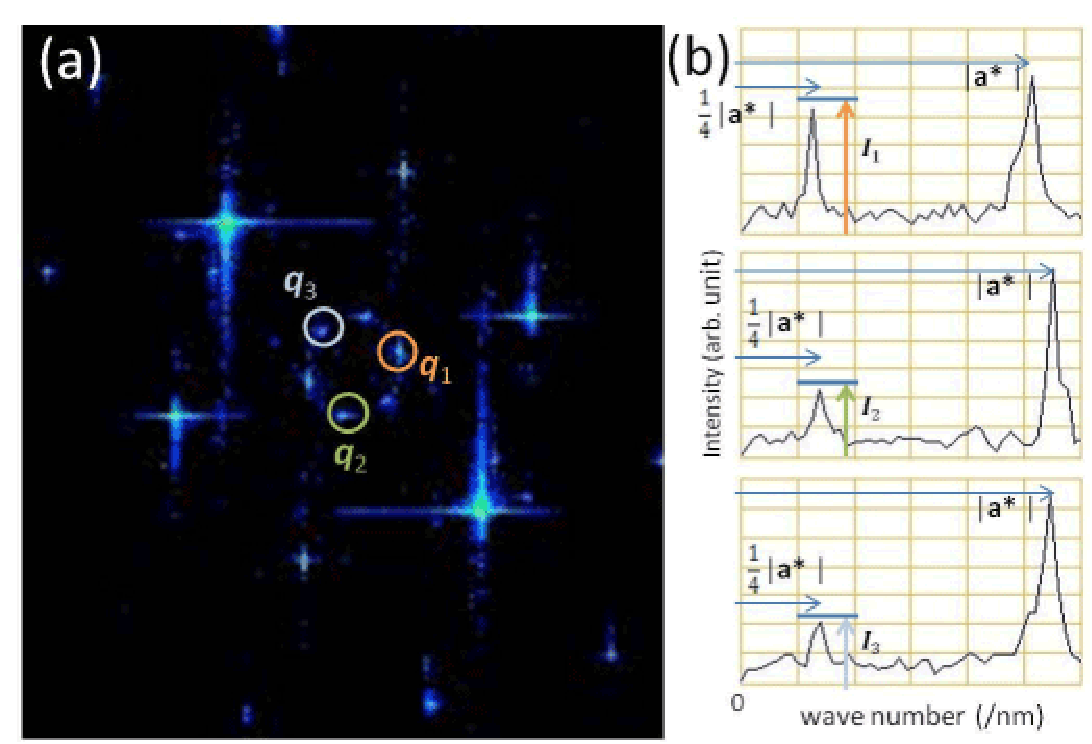}
\caption{\label{fourier}(a) Fourier transformation (FT) image of Fig. 1. 
The spots surrounded by circles show the CDW wave vectors $q_{1}$, $q_{2}$ and $q_{3}$. 
(b) Line profiles along the $q_{1}$, $q_{2}$ and $q_{3}$ wave vectors.
$I_{1}$, $I_{2}$ and $I_{3}$ show the intensities of each CDW peak. FT was performed over the entire field of view (13 nm ~ 13 nm) of the STM image.
}
\end{figure}

To investigate that periodic structure, we analyzed a Fourier transformation (FT) image of the STM.
Figure \ref{fourier}(a) shows an FT image of the entire field of Fig. \ref{stm}.
The bright spots surrounded by circles correspond to CDW vectors $q_{1}$, $q_{2}$ and $q_{3}$.
The outer intensity peaks correspond to the Bragg peaks of the selenium lattice, and 
the inner peaks correspond to the CDW satellite peaks.
Figure \ref{fourier}(b) shows line profiles of an FT image along the $q_{1}$, $q_{2}$ and $q_{3}$ wave vector.
$I_{n}$ ($n$ = 1, 2, 3) represents the intensity of CDW satellite peaks along each wave vector.
We found an obvious difference between the CDW intensities of CDW vectors $q_{1}$, $q_{2}$ and $q_{3}$; 
$I_{1}$ is the strongest and $I_{2}$ and $I_{3}$ are the second strongest and weakest, respectively.
In typical two-dimensional CDWs $I_{1}$, $I_{2}$ and $I_{3}$ must be of equal value and the CDW structure is expected to have six-fold symmetry.
However in 1\textit{T}-VSe$_{2}$ each $I_{n}$ is different, so the CDW of 1\textit{T}-VSe$_{2}$ has lower symmetry.
We concluded that is a chiral CDW by analyzing an FT image and decided that the CDW structure is clockwise 
because the direction from $q_{1}$ to $q_{2}$ to $q_{3}$ in descending order of the magnitude of $I_{n}$ is clockwise in an FT image
as reported elsewhere \cite{ishi2}.

\begin{figure}
\includegraphics{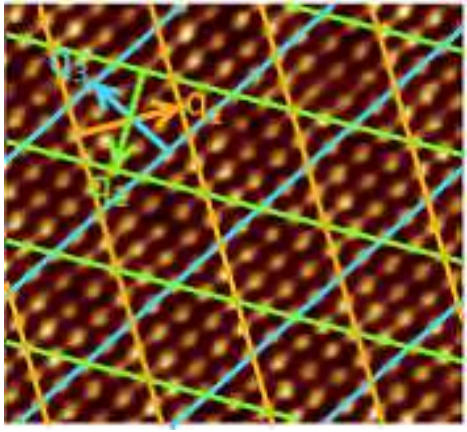}
\caption{\label{kagome}Enlarged view of a portion of Fig. 1. The orange lines show the strongest intensity of the CDW. 
The green and blue lines show the second strongest and weakest intensities of the CDW peak, respectively.
The lines form a kagome lattice.}
\end{figure}

Then we consider the CDW formation in real space from the result of an FT image.
Figure \ref{kagome} is an enlarged view of Fig. \ref{stm}.
The orange lines show the wave front of the strongest intensity CDW wave vector $q_{1}$ found in Fig. \ref{fourier}.
The green and blue lines show the second strongest and weakest intensity CDW wave fronts of $q_{2}$ and $q_{3}$, respectively.
We found that the CDW peaks form a kagome lattice rather than a triangular lattice.
Previous papers on STM observations of 1\textit{T}-VSe$_{2}$ did not report on this matter\cite{stm,stm2,stm3}.
The kagome lattice, which has an intensity difference, has two-fold symmetry.
This is unlike a conventional kagome lattice, which has three-fold symmetry and no mirror symmetry.
As a result of the lower symmetry mentioned above, we conclude that the CDW on 1\textit{T}-VSe$_{2}$ is also a chiral CDW in real space.

\begin{figure}
\includegraphics[width=12cm]{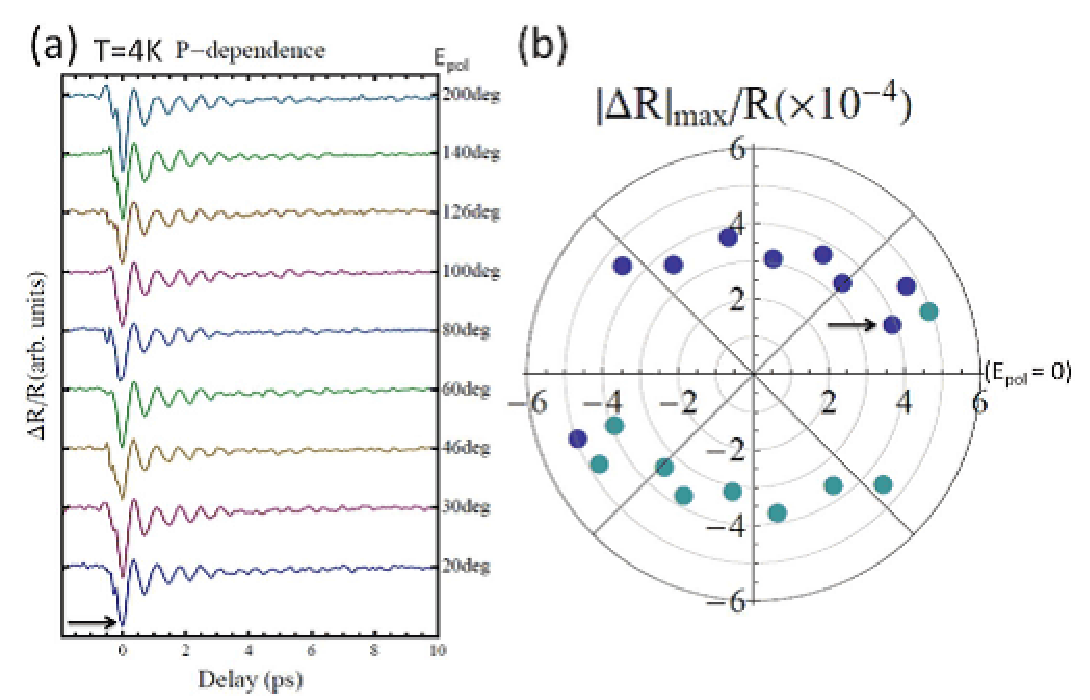}
\caption{\label{pump}(a) Polarization dependence of transient reflectivity variation in 1\textit{T}-VSe$_{2}$ measured at 4 K.
The hight of each peak is plotted in (b) as a function of polarization angle. The amplitude indicated the black arrows in (a) correspond to 
the spot indicated that in (b). }
\end{figure}

Secondly, we measured the polarization dependence of the transient reflectivity 
of 1\textit{T}-VSe$_{2}$ by using the pump-probe method to investigate the macroscopic symmetry.
With an optical measurement using a probe light whose penetration length is several tens of nanometers, we can measure the symmetry of more deeper samples 
than that with an STM. 
Figure \ref{pump} shows the polarization dependence of the transient reflectivity variation of 1\textit{T}-VSe$_{2}$ at 4 K.
The heights of the peak intensities of each polarization angle in Fig. \ref{pump} (a) are plotted in Fig. \ref{pump} (b).
The transient variation in optical reflectance $\Delta$R(t) was measured by changing the polarization angle $E\rm_{pol}$ of the incident probe laser pulse.
In general, the peak intensity of transient signals corresponds to the number of electrons excited over the CDW gap \cite{pump1,pump2}.
Figure \ref{pump} (b) shows that 1\textit{T}-VSe$_{2}$ exhibits two-fold symmetry despite the crystal structure being a triangular lattice.
For this reason, this result reflects the CDW characteristics as two-fold symmetry.
That is similar to the result with 1\textit{T}-TiSe$_{2}$ \cite{ishi1} and consistent with the result of the STM measurement on a microscopic scale.
We also confirmed that the CDW in 1\textit{T}-VSe$_{2}$ is a chiral CDW with an optical polarimetry measurement.

\begin{figure*}
\includegraphics[width=12cm]{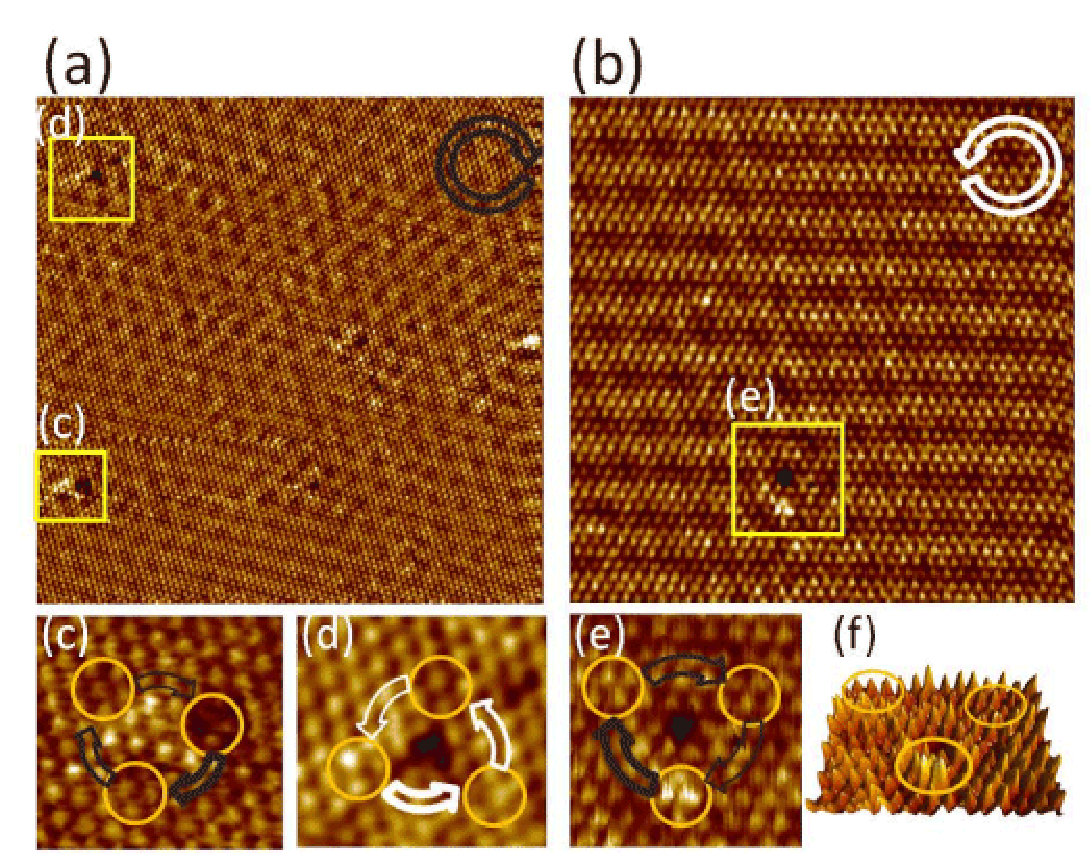}
\caption{\label{fo}(a) 24 nm ~ 24 nm STM image measured at 80 K in the clockwise CDW. (b)14 nm ~ 14 nm STM image measured at 8 K in the counterclockwise CDW. 
The Friedel oscillations in the clockwise CDW [(c) and (e)] and counterclockwise CDW (d).
Each peak and gap is indicated by orange circles. The black and the white arrows indicate clockwise and counterclockwise CDWs and FOs, respectively. 
(f)Three-dimensional plots of (e).}
\end{figure*}

Next, 
we analyzed the charge density modulation arround the charge impurities such as the lattice defects and add atoms observed in the STM measurements.
Such modulations are called Friedel Oscillations (FOs).
FOs are damped oscillations whose wave number is Fermi wave number $k\rm_{F}$, so an analysis of the FOs in CDW systems is important for 
the elucidation of CDW characteristics.
Figures. \ref{fo} show the results analyzed the current images.
The chirality of the underlying CDW is obtained by the process used in Fig. \ref{fourier}.
Figure \ref{fo}(a) is an STM image obtained in a clockwise CDW and Figure \ref{fo}(b) shows an STM image obtained in a conuterclockwise CDW. 
We observe the dark FO around a bright spot whose charge density is rich has clockwise chirality in a clockwise CDW in Fig. \ref{fo}(c), while the bright FO around a dark spot whose charge density is poor has counterclockwise chirality in the clockwise CDW in Fig. \ref{fo}(d).  
Furthermore, we found that the bright FO around the dark spot has clockwise chirality in the counterclockwise CDW.
That is, the chiralities of bright and dark FOs are the same and the opposite of the underlying CDW, respectively.
We cannot observe any opposite chirality relationship between an underlying CDW and FOs in Fig \ref{fo}(b).
In the previous study of chiral CDWs in 1\textit{T}-TiSe$_{2}$, 
dark FOs around the bright spot also have the opposite chirality and bright FOs around the dark spot also have the same chirality as the underlying CDW\cite{ishi2}.
Thus the chirality of FOs in 1\textit{T}-VSe$_{2}$ is the same as those in 1\textit{T}-TiSe$_{2}$, 
but the periodicity of FOs in 1\textit{T}-VSe$_{2}$ is longer than those in 1\textit{T}-TiSe$_{2}$
because the CDW periodicities of 1\textit{T}-VSe$_{2}$ and 1\textit{T}-TiSe$_{2}$ are 4$a_{0}$ and 2$_{0}$, respectively.
We found that the chirality of FOs reflects the underlying CDW chirality and periodicity.

\begin{figure*}
\includegraphics[width=16cm]{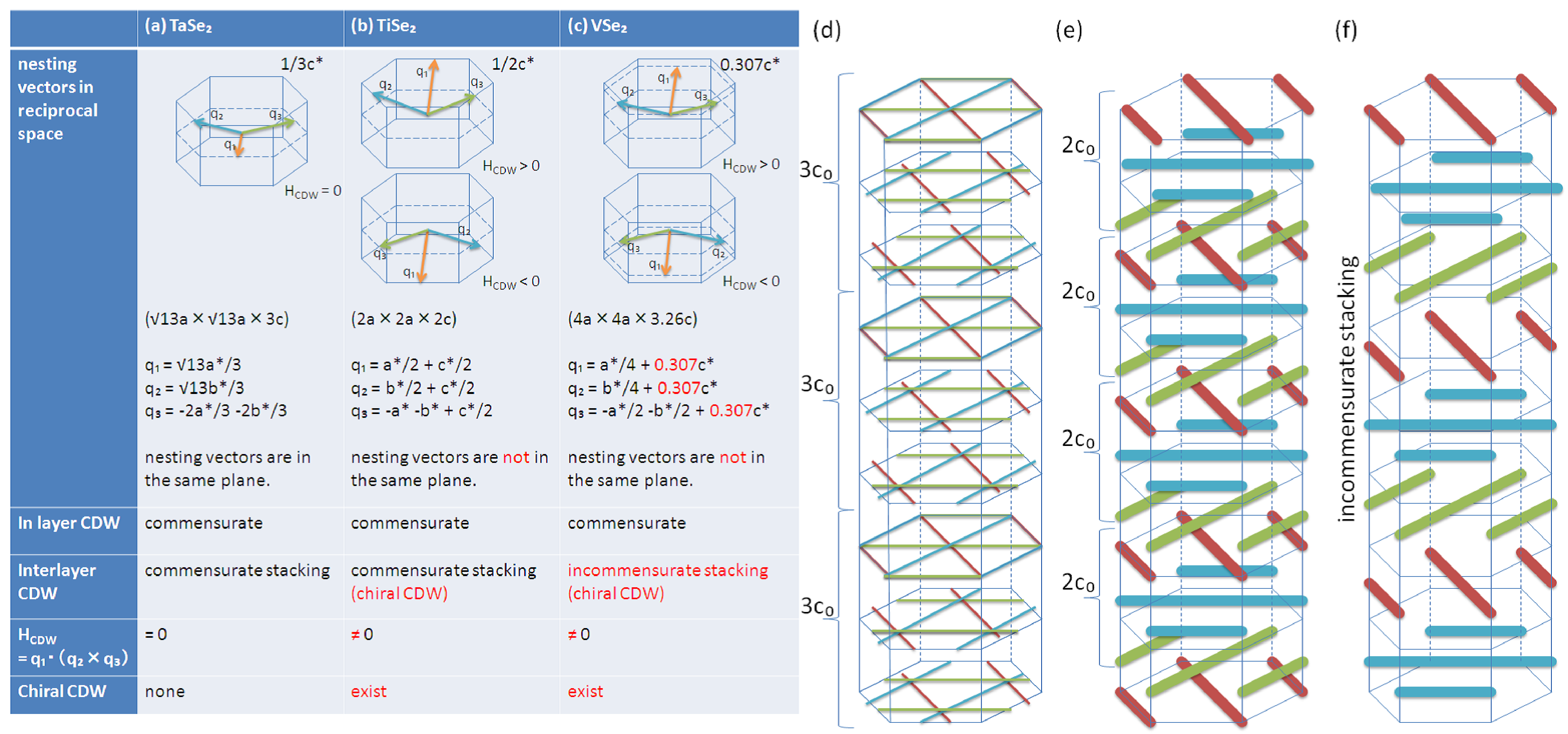}
\caption{\label{table}Table showing details of the CDW in 1\textit{T}-MSe$_{2}$ group (M = Ta, Ti, and V) [(a),(b), and (c)] 
and a schematic representation of each material [(d), (e), and (f)]. 
We provide pictures of nesting vectors in reciprocal space, 
the values of each nesting vector, the commensurateness of layer and interlayer CDWs, 
the value of $H_\mathrm{CDW} \equiv q_{1}\cdot(q_{2} \times q_{3})$ and the exsistence of the chiral CDWs of (a) 1\textit{T}-TaSe$_{2}$, (b) 1\textit{T}-TiSe$_{2}$ and 
(c) 1\textit{T}-VSe$_{2}$.
(a) 1\textit{T}-TaSe$_{2}$ is a typical two-dimensional CDW material and its nesting vectors are in the same plane, 
so $H_\mathrm{CDW} = 0$. The superlattice is 3$c_{0}$, which originated from the stacking. There is no chiral CDW. 
(b) The nesting vectors of 1\textit{T}-TiSe$_{2}$ are not in the same plane, so $H_\mathrm{CDW} \neq 0$. Its interlayer CDW is commensurate. 
There is a chiral CDW. 
(c) The nesting vectors of 1\textit{T}-VSe$_{2}$ are not in the same plane, so $H_\mathrm{CDW} \neq 0$. Its interlayer CDW is incommensurate. 
There is a chiral CDW. 
Schematic representations of (d) a typical two-dimensional CDW similar to 1\textit{T}-TaSe$_{2}$ and chiral CDW in (e) 1\textit{T}-TiSe$_{2}$ and (f) 1\textit{T}-VSe$_{2}$.
Each hexagon shows a layer of each material. Colored lines show the charge concentration and the colors correspond to the CDW $q$ vectors in (a), (b) and (c).
}
\end{figure*}

Finally, we discuss the condition of the chiral CDW.
Figure \ref{table} shows the details of a CDW in the 1\textit{T}-MSe$_{2}$ group (M = Ta, Ti, V).
These CDWs have the same characteristics as a CDW with commensurate in a layer but they are quite different in the $c$* direction.
The CDW $c$* superlattice of 1\textit{T}-TaSe$_{2}$ is 3$c_{0}$, which originates from CDW stacking, and this structure does not relate to the direction of the nesting vectors.
Most triple-$q$ CDW systems such as 1\textit{T}-TaS$_{2}$ have a similar structure to 1\textit{T}-TaSe$_{2}$.
In 1\textit{T}-TiSe$_{2}$, the $c$* component is comensurate as with 1\textit{T}-TaSe$_{2}$, 
but it forms a 2$c_{0}$ superlattice originating from nesting vectors with a $c$*/2 component.
This 2$c_{0}$ superlattice stacks in the $c$* direction and forms a chiral CDW.
Meanwhile, the 1\textit{T}-VSe$_{2}$ CDW is unique in triple-$q$ CDW systems as it is incommensurate along the $c$* direction 
because the $c$* component of its nesting vectors is an irrational number \cite{xray2}.
We cannot consider 1\textit{T}-VSe$_{2}$ as composing the stacking unit, 
so the chiral CDW structure would appear directly below the transition temperature.
In this respect, 1\textit{T}-VSe$_{2}$ is a more important chiral CDW material than 1\textit{T}-TiSe$_{2}$.
A chiral CDW was discovered in both 1\textit{T}-TiSe$_{2}$ and 1\textit{T}-VSe$_{2}$ but there is little difference between them.
Nevertheless, $H_\mathrm{CDW} \neq0$ is probably a necessary condition of the chiral CDW.

\section{CONCLUSIONS}
We discovered a chiral CDW in 1\textit{T}-VSe$_{2}$ by using STM measurements and optical polarimetry measurements and by analyzing of the FOs.
We also discovered that the CDW peaks in 1\textit{T}-VSe$_{2}$ form a kagome lattice.
1\textit{T}-VSe$_{2}$ has a unique characteristic as an incommensurate CDW along the $c$* direction, which is different from other MX$_{2}$ triple-$q$ CDWs, 
but it can be treated is the same way as 1\textit{T}-TiSe$_{2}$ in terms of $H_\mathrm{CDW} \equiv q_{1}\cdot(q_{2} \times q_{3}) \neq0$.
These results strongly suggest that the chiral CDW is a general phenomenon that can occur in systems that fulfill the condition $H_\mathrm{CDW}\neq0$.



\nocite{*}

\begin{thebibliography}{9}
\bibitem{chiraldw}I. E. Frolov, V. Ch. Zhukovsky, and K. G. Klimenko, 
Phys. Rev. D \textbf{82},076002(2010)
\bibitem{texture}P. Poushan, J. Seo, C. V. Parker, Y. S. Hor, D. Hsieh, D. Qian, A. Richardella, M. Z. Hasan, R. J. Cava, and A. Yazdani, 
Nature \textbf{460},1106(2009)
\bibitem{emcha1}G. L. J. A. Rikken, J. F\"{o}lling, and P. Wyder, 
Phys. Rev. Lett. \textbf{87},236602(2001)
\bibitem{emcha2}V. Krsti\.{c}, S. Roth, M. Burghard, K. Kern and G. L. J. A. Rikken, 
J. Chem. Phys. \textbf{117},11315(2002)
\bibitem{emcha3}F. Pop, P. Auban-Senzier, E. Canadell, G. L. J. A. Rikken, and N. Avarvari, 
Nat. Commun. \textbf{5},3757(2014)
\bibitem{pwave}H. Nobukane, A. Tokumo, T. Matsuyama, and S. Tanda,
Phys. Rev. B \textbf{83},144502(2011)
\bibitem{hightc1}P. Hosur, A. Kapitulnik, S. A. Kivelson, J. Orenstein, and S. Raghu, 
Phys. Rev. B \textbf{87},115116(2013)
\bibitem{hightc2}S. S. Pershoguba, K. Kechedzhi, and V. M. Yakovenko, 
Phys. Rev. Lett. \textbf{111},047005(20013)
\bibitem{ishi1}J. Ishioka, Y. H. Liu, T. Kurosawa, K. Ichimura, Y. Toda, M. Oda, and S. Tanda, 
Phys. Rev. Lett. \textbf{105},176401(2010)
\bibitem{ishi2}J. Ishioka, T. Fujii, K. Katono, K. Ichimura, T. Kurosawa, M. Oda, and S. Tanda, 
Phys. Rev. B \textbf{84},245125(2011)
\bibitem{MX2}J. A. Wilson, F. J. Di Salvo, and S. Mahajan, 
Phys. Rev. Lett. \textbf{32},882(1974)
\bibitem{2h}I. Guillam\'{o}n, H. Suderow, J. G. Rodrigo, S. Vieira, P. Rodi\'{e}re, L. Cario, E. Navarro-Moratalla, C. Mart\'{i}-Gastaldo, and E. Coronado,
New J. phys. \textbf{13},103020(2010)
\bibitem{xray}K. Tsutsumi, T. Sambongi, A. Toriumi, and S. Tanaka, 
J. Phys. Soc. Jpn. \textbf{49},2(1980)
\bibitem{xray2}K. Tsutsumi,
Phys. Rev. B \textbf{26},10(1982)
\bibitem{band}A. Zunger et al, 
Phys. Rev. B\textbf{19},6001(1979)
\bibitem{arpes} T. Sato, K. Terashima, S. Souma, H. Matsui, T. Takahashi, H. Yang, S. Wang, H. Ding, N. Maeda, and K. Hayashi, 
J. Phys. Soc. Jpn. \textbf{73},3331(2004)
\bibitem{arpes2} V. N. Strocov et al, 
Phys. Rev. Lett. \textbf{109},086401(2012)
\bibitem{vse3}A. V. Skripov, D. S. sibirtsev, Yu G. Cherepanov, and B. A. Aleksashin, 
J. Phys.: Condens Matter \textbf{7},4479(1995)
\bibitem{stm} R. V. Coleman, B. giambattista, P. K. Hansma, A. Johnson, W. W. McNairy and C. G. Slough, 
Adv. Phys. \textbf{37},559(1988)
\bibitem{stm2} B. giambattista, C. G. Slough, W. W. McNairy, and R. V. Coleman, 
Phys. Rev. B \textbf{41},10082(1990)
\bibitem{stm3} J. J. Kim, C. Park, and H. Olin, 
J. Korean Phys. Soc. \textbf{31},713(1997)
\bibitem{pump1}J. Demsar, K. Biljakovic, and D. Mihailovic, 
Phys. Rev. Lett. \textbf{83},800(1999)
\bibitem{pump2}J. Demsar, L. Forro, H. Berger, and D. Mihailovic, 
Phys. Rev. B. \textbf{66},041101(2002)
\end{thebibliography}
\end{document}